# Label-free, *in situ* monitoring of viscoelastic properties of cellular monolayers via elastohydrodynamic phenomena

Tianzheng Guo,[1] Xiaoyu Zou,[1] Shalini Sundar,[2] Xinqiao Jia,[1,2] Charles Dhong[1,2]*

[1]Department of Materials Science and Engineering, University of Delaware, Newark, Delaware, USA.
[2]Department of Biomedical Engineering, University of Delaware, Newark, Delaware, USA.
*Corresponding author. Email: cdhong@udel.edu

**ABSTRACT:**
Recent advances recognize that the viscoelastic properties of epithelial structures play important roles in biology and disease modeling. However, accessing the viscoelastic properties of multicellular structures in mechanistic or drug-screening applications face challenges in repeatability, accuracy, and practical implementation. Here, we present a microfluidic platform that leverages elastohydrodynamic phenomena, sensed by graphene strain sensors, to measure the viscoelasticity of cellular monolayer *in situ*, without using labels or specialized equipment. We demonstrate platform utility with two systems: cell dissociation following trypsinization, where viscoelastic properties change over minutes, and an epithelial-to-mesenchymal transition, where changes occur over days. These cellular events could only be resolved with our platform's higher signal-to-noise ratio: relaxation times of $14.5 \pm 0.4$ s$^{-1}$ for intact epithelial monolayers versus $13.4 \pm 15.0$ s$^{-1}$ in other platforms. By rapidly assessing combined contributions from cell stiffness and intercellular interactions, we anticipate that the platform will hasten translation of new mechanical biomarkers.

**INTRODUCTION:**
Recent work in fundamental biology has increasingly developed connections between cell and tissue stiffness with function or state (*1, 2*). While several techniques can assess single cell mechanical properties, there are few methods that can extract cell monolayer properties. Cell monolayers, in addition to individual cell stiffness, contain additional mechanical contributions from cell-cell and cell-protein interactions.(*3*) Cells collectively interact with neighboring cells and surrounding proteins inside cell monolayers, and cell monolayers play a critical role in many physiological functions (*4*). In particular, the stiffness of cellular monolayers is both directly correlated to function, like in maintaining structural integrity, and indirectly, like in modulating mechanotransduction to downstream cellular pathways (*5, 6*). Aberrant monolayer stiffness is known to indicate disease states (*7*), from age-related macular degeneration in visual impairment (increased stiffness) (*8*), in degradation of epithelial layers in intestinal inflammatory diseases (decreased stiffness) (*9*), to scarring in fibrosis (increased stiffness) (*10*). These insights from biology position the stiffness of cellular monolayers as a versatile and novel biomarker for pathogenesis or pathophysiology, but a key limitation to translation has been the lack of simple platforms which can rapidly and accurately measure changes in the stiffness of cellular monolayers.

Ideally, a platform for measuring monolayer stiffness should have the following features (*11–13*): it should give accurate and repeatable measurements, integrate into current cell culturing workflows, and measure stiffness *in situ* instead of only at endpoints. It should also avoid complicating experiments by introducing new labels or reagents, and not require capital-intensive or specialized equipment. Current techniques that assess the stiffness of cellular monolayers are limited by one or more of these burdens, which ultimately reduce the adoption and translation of mechanical stiffness as a biomarker (*14–16*). For example, a common technique is transepithelial/transendothelial electrical resistance (TEER) (*17*). TEER quantitatively measures the tight junction dynamics in cell culture models of endothelial or epithelial monolayers by measuring how much of the electrical signal is blocked by the cell layer. Experimental constraints specific to TEER make it difficult to conduct biochemical assays on the same specimens, which increases experimental complexity. Furthermore, without specific expertise, TEER may be functionally limited to a binary output of either successful or unsuccessful monolayer formation, and reproducibility remains a challenge. Similarly, optical techniques such as optical tweezers (*18*) or traction force microscopy (*19*) require significant user knowledge to operate specialized equipment, develop appropriate controls, and perform analysis.

In other indirect techniques which translate non-mechanical events to estimate mechanical parameters, Li et al. (*20*) characterized the viscoelastic properties of fibroblast cell monolayers using thickness-shear-mode acoustic wave sensors. The cell layer was cultured onto the quartz resonator substrate to obtain the electrical admittance spectrum which would reflect acoustic impedance change caused by cell monolayer adhesion. However, an interfacial layer between cell monolayer and substrate with unknown thickness would cause variations, and the calibration process is complex. Adamo et al. (*21*) developed a microfluidic assay by measuring changes in cellular shape and transit times as cells pass through constrictions to obtain cell stiffness. Similarly, Jochen et al. (*22*) used a microfluidic optical stretcher to deform cells in order to find a link between cell function and elasticity. However, the refractive index of each sample needs to be determined before stretching which increases the operative difficulty. Both microfluidic approaches achieved a decreased measurement time, which improved throughput, but the techniques still require the assistance of capital-intensive optical-based characterization. Most existing techniques are designed for single cell assessments(*23, 24*) and thus neglect the cell-cell and cell-protein interactions, which are important in monolayers or multicellular systems.

Here, to overcome existing limitations with current measurement techniques, we present a microfluidic-based platform which relies on the theory of elastohydrodynamic deformation (EHD)—the deformation of elastic walls under flow—as measured through piezoresistive graphene strain sensors (*25*). This platform capable of non-invasive stiffness measurements *in situ*, alongside conventional cell culturing conditions, without the need for optical components or new reagents. By relying on relatively simple voltage measurements as opposed to microscopy techniques, the technique has better scaling for throughput or parallel experiments.

We demonstrate the platform utility through cellular monolayer dissociation via trypsinization, which occurs in minutes, as well as stiffness changes induced by a morphology change from an epithelial-to-mesenchymal transition (EMT), which occurs over days. Our microfluidic-based platform provides a new facile route for investigating dynamic cellular mechanics at the multicellular level.

**RESULTS**
**Operational principle and device validation**
Our platform is based on monitoring the flow-induced deformation of soft walls (elastohydrodynamic deformation, EHD) within a microfluidic channel (*26–29*). To extract the viscoelastic parameters of a sample, we monitored the viscoelastic relaxation of the wall to its baseline shape after perturbation by a droplet (Schematic shown in **Fig. 1A**). For a pure fluid, Gervais et al. (*30*) developed a model to describe the expected wall deformation as a function of parameters. Given the addition of a sample, i.e., monolayer of cells, inside the fluid channel, the elastohydrodynamic deformation of the wall under flow will change in a manner which reflects the mechanical properties of the sample. Under a constant flow rate, the relationship between the flow rate and the pressure and the channel deformation are given by (*31*):

$$Q = \frac{h_0^4 E}{48 c_2 \mu (L-z)} \left[ \left(1 + c_2 \frac{p(z)W}{E h_0}\right)^4 - 1 \right] \quad (1)$$

$$\frac{\Delta h}{h_0} = 2 \sqrt[4]{\frac{3 Q \mu (L-z)}{E h_0^4} + 1} - 2 \quad (2)$$

where $Q$ is the flow rate, $h_0$ is the initial channel height, $E$ is the elastic modulus of the channel walls in the device, $L$ is the length of the channel, $z$ is the axial position along the channel, $\mu$ is the fluid viscosity, $W$ is the width of the channel, $p(z)$ is the pressure distribution along the channel. The addition of a viscoelastic layer alters the viscoelastic relaxation of the channel by changing layer thickness and elastic modulus. In principle, by fitting a multilayered Kelvin-Voigt model to the deformation versus time during the relaxation phase, viscoelastic parameters of each layer (the properties of the channel are known, the sample is unknown) are obtained. Although theoretical formulations exist (*32–36*), due to the finite geometry of our microfluidic channel, we also used finite element analysis (FEA) modeling to extract the geometric-dependent contributions from the material properties in the relaxation profile of our microfluidic channel (See **Supplementary Materials**, **Fig. S1**).

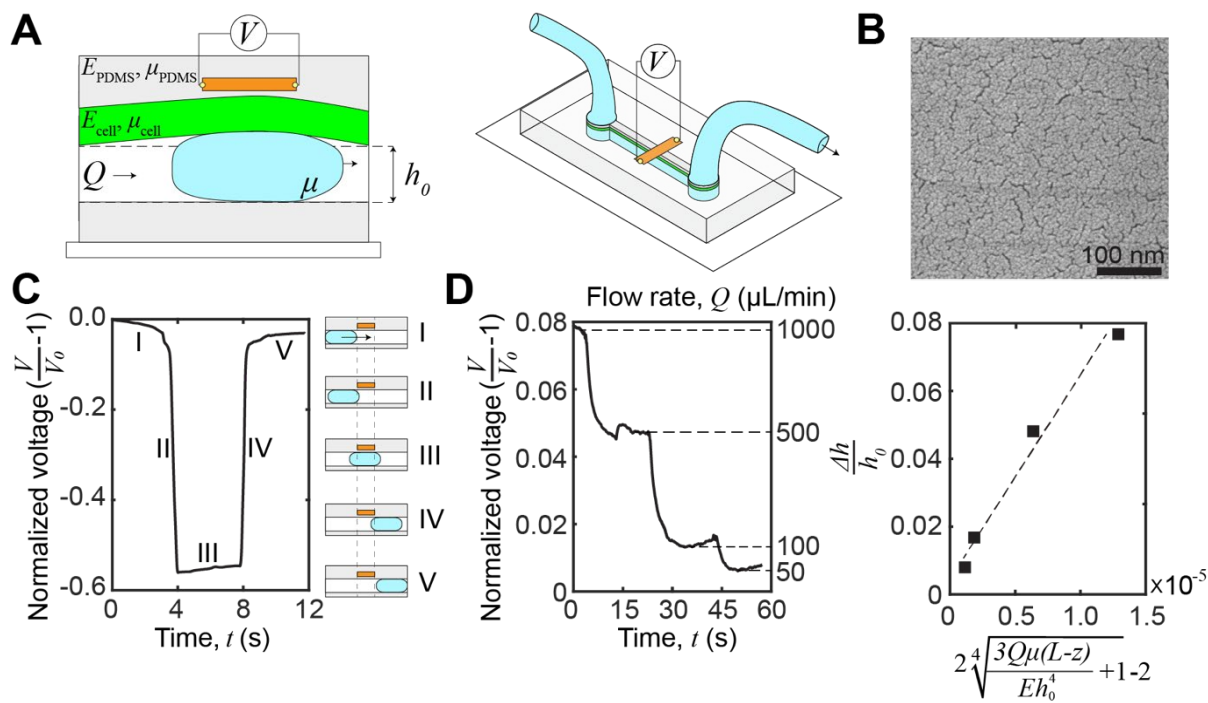

**Fig. 1. Microfluidic device and operation principle.** (**A**) Schematic of the microfluidic device. A graphene strain sensor, embedded in the PDMS sidewalls, is placed in close proximity (~8-10 μm) to the channel. The sensor measures the transient deformation caused by a passing droplet, and this deformation is a function of the properties of the channel sidewalls and any coatings, i.e., the cellular monolayer. (**B**) Scanning electron microscope (SEM) images of palladium nanoislands on graphene. (**C**) Voltage signal capture by introducing a water droplet in the channel. The voltage signal can be broken into distinctive regions which correspond to the location of the droplet relative to the channel and sensor. Region IV is used for relaxation analysis. (**D**) Device calibration by monitoring the voltage change of the device at different flow rates. Experimental data fitted against theoretical prediction of deformation by Eq. (2). The relationship between flow rate and elastohydrodynamic deformation is approximately linear at low pressures.

The expected deformation in most convention microfluidic devices are low (< 5 μm) (*37*), and the relaxation times are fast (~0.1 s) (*31*). To measure this deformation, we embedded a graphene strain sensor within the device sidewalls (**Fig. 1A** and **Fig. 5**), in close proximity (~8-10 μm) to the channel (*31*). These piezoresistive strain sensors are made from graphene decorated with metallic nanoislands (**Fig. 1B**), which have sufficient sensitivity to measure elastohydrodynamic phenomena (*25*). The advantages of using a piezoresistive strain sensors is that the electrical signal monitoring has an inherently high temporal resolution and a lower cost for scale up, especially when compared to optical-based lasers, cameras, or microscope-based platforms. Furthermore, by embedding the sensor within the elastic wall, the sensor reduces erroneous sources of contamination and does not introduce additional experimental complications, e.g., tracer particles or conjugated labels. These advantages enable widespread compatibility, scalability, and sufficient mechanical resolution for changes at the cellular level.

To induce a transient deformation in the microfluidic channel, we flowed a water droplet into the channel. The voltage readings of the water droplet can be broken into five distinctive regions (**Fig. 1C**). In region I, we introduced a droplet into the channel, but as the droplet was far away from the sensor, the voltage showed a slight decrease which corresponds to the applied pressure driving the droplet, constricting the channel ahead of the droplet (*38*). In region II, the leading edge of the droplet crosses the boundary (dash line on the left in **Fig. 1C**) of the sensor, leading to a reduction in voltage. Thus, we can calculate the droplet velocity by dividing the width of the strain sensor by the duration of region II. In region III, the main body of the droplet transits across the sensor strip, which results in a slight increase in the voltage due to the increasing area of the overlap between the droplet and the sensor. In region IV, the droplet has left the sensing area, and the voltage signal reflects the viscoelastic relaxation of the channel. Region IV is where the voltage data can be fitted to a viscoelastic model of the sample to extract viscoelastic parameters of the cellular monolayer. (See **Methods** for fitting procedure). Region V shows a second, slower relaxation of the channel back to the neutral height. This relaxation is a bulk phenomenon and is distinct from the interfacial viscoelastic relaxation in region IV. The bulk residual strain of the channel originates from the confined microfluidic geometry, as replicated in our simulations (see **Supplementary Materials, Fig. S1**).

To obtain a calibration for the deformation response of the sensor, we used the theoretical relationship between flow rate and channel deformation in Eq. (2). We performed calibration *in situ* by using a PID-controlled microfluidic controller (Elveflow OB1 MK3+, 61 μbar resolution) to vary the flow rate while monitoring the voltage change. At lower pressures, the pressure drop is approximately linear with the imposed flow rate (*30*). **Fig. 1D** shows that the relationship between channel deformation and flow rates is as expected, demonstrating that strain sensor is sensing elastohydrodynamic deformation and that the voltage readings are proportional to channel deformation.

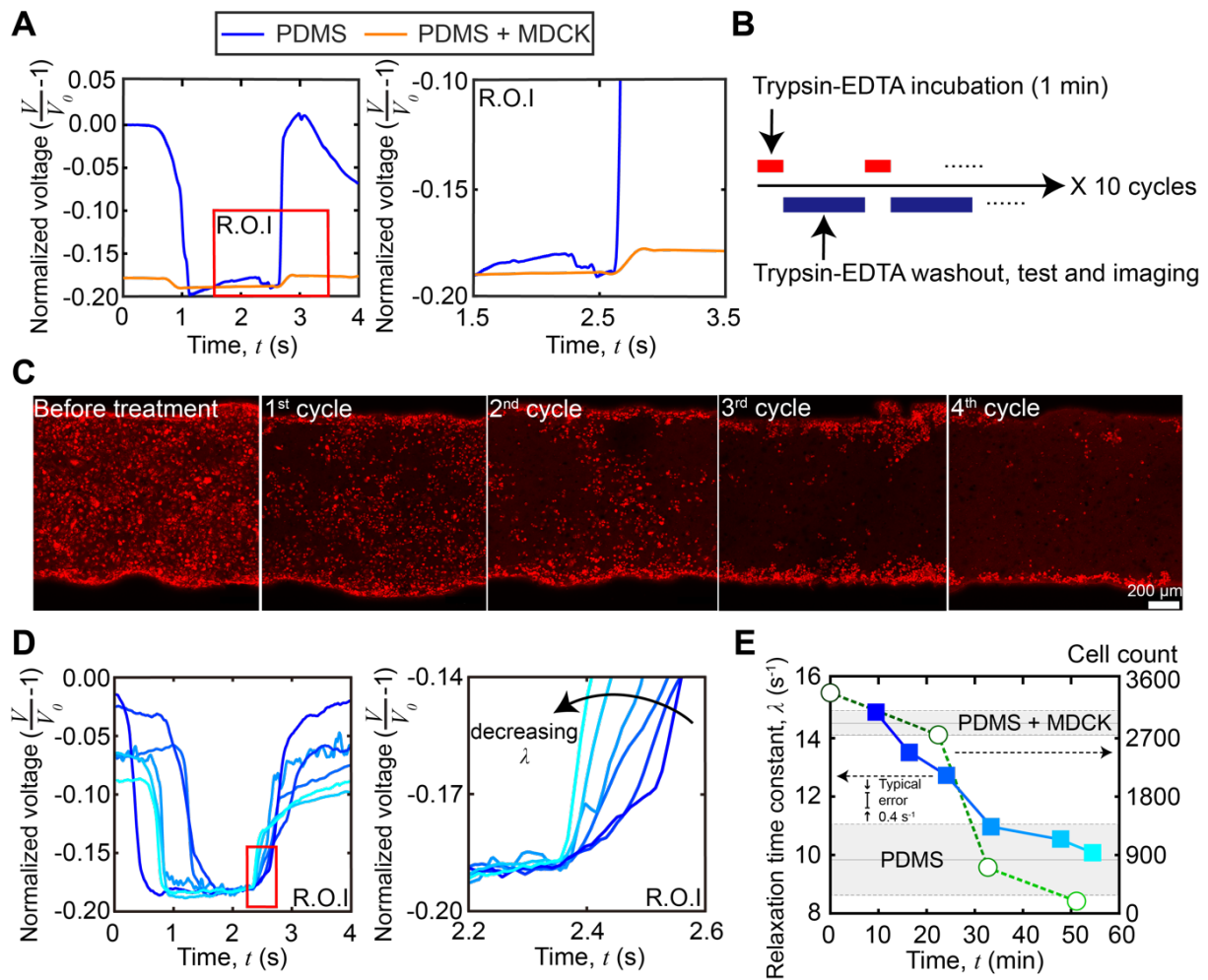

**Fig. 2.** *In situ* **stiffness measurement of trypsinization of the MDCK cell monolayer.** (**A**) The raw sensor data of bare PDMS and MDCK cells coated PDMS. Signals are overlayed to aid visual comparisons of viscoelastic relaxation between experimental conditions. (**B**) Trypsinization treatment sequence. Trypsin-EDTA incubation and the following washout, test and imaging are considered one cycle. The complete experimental monitoring contains 10 cycles. (**C**) Confocal microscopy images of MDCK cells in the microchannel during trypsinization (cells were labeled with CellTracker red, Life Technologies). After the 4th cycle, the cell number drops to below 500, i.e., the threshold indicating complete cellular detachment. Scale bar = 200 μm.(**D**) The raw viscoelastic relaxation data of different time points of trypsinization treatment. Signals are overlayed to aid visual comparisons of viscoelastic relaxation, and the blue intensity indicates time points (**E**) Relaxation time constant overlayed with change in cell count as a function of time during trypsinization. The typical single measurement error from technical replicates is 0.4 s$^{-1}$.

**Transient viscoelastic properties of a monolayer under trypsinization.**

To demonstrate *in situ* measurements of dynamic stiffness changes in cellular monolayers, we dissociated a confluent Madin-Darby canine kidney (MDCK) epithelial cell monolayer with a standard 0.25% Trypsin-EDTA (ethylenediaminetetraacetic acid) solution (See **Methods** for channel coating, culture conditions, and determination of confluency). Prior to trypsinization, we obtained baseline viscoelastic relaxation time constants through of PDMS ($\lambda = 9.9 \pm 1.3$ s$^{-1}$, $N = 10$, no significant difference with and without collagen IV coating on the channel wall) and a fully formed MDCK cell monolayer ($\lambda = 14.5 \pm 0.4$ s$^{-1}$, $n = 3$, $N = 10$). The higher $\lambda$ value obtained in the channel with the MDCK monolayer indicates a higher overall stiffness within the channel. That is, although the MDCK cell monolayer (8 μm) is not necessarily stiffer than the PDMS (~8-10 μm), the composite structure of PDMS with MDCK cell monolayer is stiffer than the thinner PDMS structure alone. The mechanical contributions of MDCK cell monolayer originate from a combination of the MDCK cell body, intercellular adhesion junctions, and extracellular matrix proteins (ECM) (*39*). The value are in agreement with the previous study (*40*) of $13.4 \pm 15.0$ s$^{-1}$ but have much lower error, which are important for resolving small mechanical changes during trypsinization. The trypsinization design is shown in **Fig. 2B**, which alternates between Trypsin-EDTA incubation and measurement acquisition stages. Confocal microscope images of the monolayers were taken (cells were labeled by CellTracker red, Life Technologies, **Fig. 2C**) concurrently with our viscoelastic measurements, which shows that the orthogonal nature of our stiffness measurements enables standard use of fluorescent assays. (*n*: biological replicates. *N*: technical replicates)

The raw viscoelastic relaxation data of the different time points are shown in **Fig. 2D**. When plotted versus time, **Fig. 2E** shows the relaxation time constant change over time as compared to the change in cell count during the trypsinization process. (For clarity, cell count and viscoelastic properties of a single experiment are shown in **Fig. 2E**, but additional replicates are in the **Supplementary Materials**, **Fig. S2**) At the earliest timepoints ($t = 0$-10 minutes, cycles 1-2), the relaxation time constants ($\lambda = 14.5 \pm 0.4$ s$^{-1}$, $n = 3$, $N = 10$) matched the baseline of confluent and intact MDCK cell monolayer inside the channel. At later timepoints, as expected, the trypsinization began to dissociate the MDCK monolayer (*41*), resulting in a decreased $\lambda$ which is below 14.5 s$^{-1}$ at around $t = 10$-20 minutes. However, when compared to the cell count, we observed that the layer softening occurred shortly after observable decreases in cell count. Continued Trypsin-EDTA solution exposure led to further decrease of relaxation time constant, until the value reached that of a bare PDMS channel ($\lambda = 9.9 \pm 1.3$ s$^{-1}$, $n = 3$, $N = 10$). Comparing the mechanical data with the cell count, the $\lambda$ of the channel remains above the baseline of PDMS values between 30-40 minutes for a short time, even after the cell count drops to effectively zero (< 500). This is likely due to residual basement membrane proteins, indicating that the platform can sense contributions from the cells and extracellular proteins on the overall mechanical stiffness of the monolayer.

In this trypsinization experiment, the loss of cells and the digestion of the ECM led to an expected decrease in stiffness. Critically, the higher signal-to-noise ratio of our platform was able to resolve the small mechanical changes occurring during the trypsinization process. The concurrent stiffness and cell count measurements captured the different stages and the contribution of cell body, cell-cell adhesion, basement membrane proteins secreted by MDCK cells, and ECM proteins in dissociation of the whole cell layer. At earlier time points, cell loss preceded the loss in stiffness, while at later time points, residual ECM and basement membrane proteins had a small, but detectable contribution to the channel stiffness where $\lambda$ is ~1.0 s$^{-1}$ higher than the PDMS baseline, even after the cells were virtually absent after 30 minutes. Compared with the rate of decrease in the relaxation time constant ($\lambda$), the rate of cell count loss is higher. However, compared to the earlier decrease of relaxation time constant where $\lambda$ decrease ~80% in 30 minutes—which is mostly due to cell count loss—the decrease of relaxation time in the last 20 minutes ($\lambda$ reduced ~20%) was more gradual when there were virtually no cells. This suggests that MDCK cell bodies contribute more to the monolayer stiffness than the ECM and basement membrane proteins. Our results are in agreement with those by Sorba et al. (*42*), who showed that in MDCK monolayers, the cell covered region had higher relaxation time constant (10.09 ± 1.42 s$^{-1}$) than the bare PDMS region (3.18 ± 1.47 s$^{-1}$), via tensile testing, and that EDTA treated MDCK cell had lower Young's modulus (6.86 ± 3.27 kPa) than untreated cells (23.3 ± 6.3 kPa).

**Thin-film correction factor for viscoelastic properties.**
In thin films, like the monolayers here, the apparent viscoelastic properties are a function of the cell thickness. We performed FEA simulations to determine the thin-film correction factor to isolate the true (thickness-independent) viscoelastic properties of the MDCK cell monolayer from contributions arising from thickness and the underlying PDMS (**Fig. 3A**). From our simulations, we obtained a simple correction factor to calculate the relaxation time constant relationship in a two-layer structure (**Fig. 3B**):

$$\lambda_{total} = \lambda_{PDMS} + \lambda_{cell} \times \exp\left(1 - \frac{a}{h}\right) \tag{3}$$

where *a* is a constant and *h* is the monolayer thickness (*43*). In our geometries, *a* is determined to be 9.4 μm and $\lambda_{cell}$ is 9.0 s$^{-1}$. This correction factor applies to cell thicknesses ranging from ~10-20 μm and the relaxation time constant increases with the layer thickness. Typically, *a* may be related to the effective contact radius of the transiting droplet on the cellular monolayer and is the same order of magnitude as the deformation into the cell layer. However, the simulation was in 2D, and from the displacement plot of cell surface, we observed that the deformation of the cell layer itself is less than 16% of its thickness, so it is possible that *a* should only be considered as a fitting parameter. Overall, while in many experiments the cell thickness remains similar and comparative viscoelastic values could be used without any correction factor, in cases where the cell thickness changes significantly, Eq. (3) provides a correction factor for the true viscoelastic properties after obtaining the cell thickness through standard confocal microscopy.

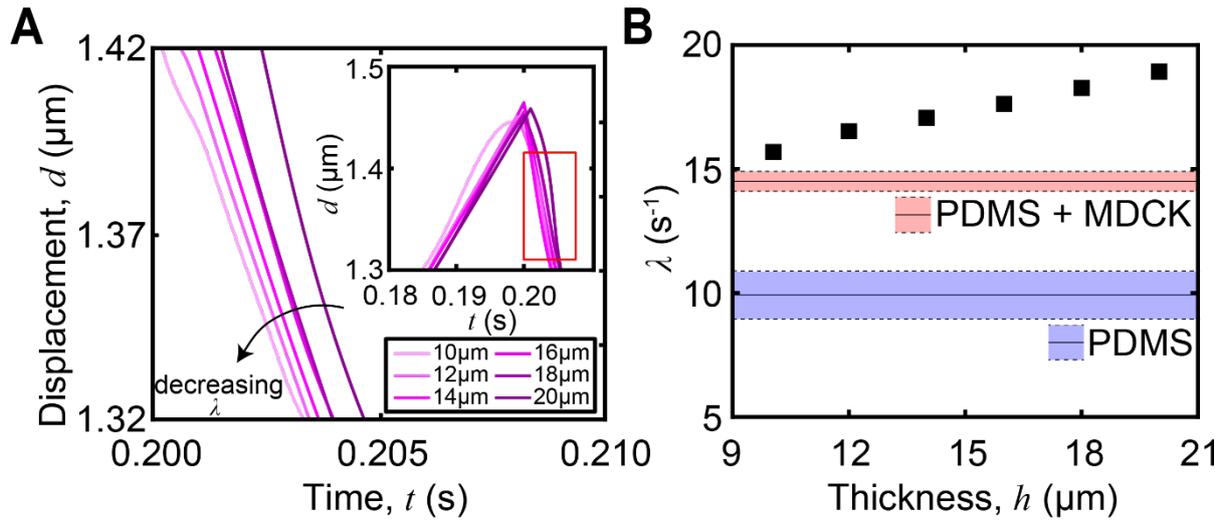

**Fig. 3. FEA simulation with varying cell layer thickness.** (**A**) The viscoelastic data from FEA simulation with cell layer thickness ranging from 10 μm to 20 μm. The relaxation section shows difference between layer thickness. Purple intensity indicates layer thickness. (**B**) Extracted $\lambda$ from the viscoelastic data of each layer thickness. It is shown that $\lambda$ increases with the layer thickness. The red and blue box indicate the baseline of the PDMS and PDMS + MDCK, respectively.

**Transient viscoelastic properties during an epithelial-to-mesenchymal transition.**

To demonstrate *in situ* stiffness changes derived solely from cellular processes like morphological changes—which does not result in cell loss—we added an EMT inducer, HGF, in the MDCK culture. EMT is a common disease marker, especially in fibrosis, which results in several biochemical, biophysical and mechanical changes (*44*, *45*). As this transition occurs slower than the trypsinization, we maintained MDCK cells in an incubator at 37 °C and with 5% $CO_2$ and 95% air. The only exception to culturing conditions are when we briefly took cells into biosafety cabinets (room temperature, ambient conditions, ~2 minutes) to obtain stiffness measurements.

To investigate the stiffness change due to EMT, we simulated an in vitro EMT process in the microfluidic channel by introduction of hepatocyte growth factor (HGF, Invitrogen) (*46*). We treated one group of MDCK cells with HGF while no HGF treatment on the control group on day 1. All other culture conditions were the same (*47*). On day 3, we removed cell culture media and flowed a droplet of PBS solution to acquire the voltage signal (**Fig. 4B**).

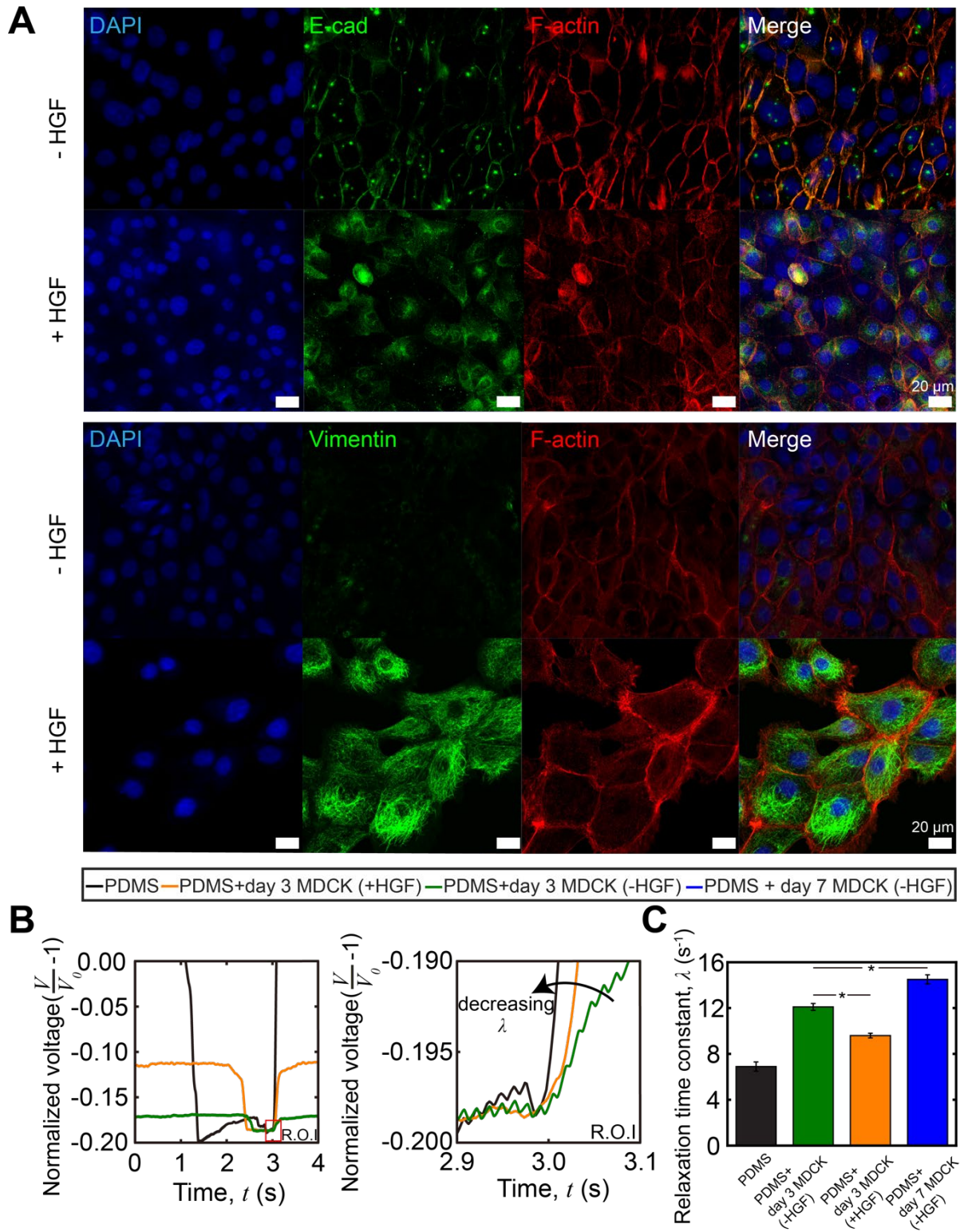

**Fig. 4. Epithelial-to-mesenchymal transition induced by hepatocyte growth factor.** (**A**) Representative immunostaining images of untreated and HGF treated MDCK cells inside the channel on day 3 showing DAPI, E-cad, F-actin and vimentin. Change of E-cad and vimentin expression indicates EMT. Scale bar = 20 μm. (**B**) The raw voltage data from the strain sensor of different treatments on the cell layer. Signals are overlayed to aid visual comparisons of viscoelastic relaxation. (**C**) Relaxation time constant comparison between untreated and HGF-

treated MDCK cells. HGF treated group shows lower value of $\lambda$, indicating a softer cell layer. MDCK cells cultured for 7 days shows a higher relaxation time constant than MDCK cells cultured for 3 days. Statistical analysis was performed using Student's t test. All quantitative analyses were conducted on data sets in which N ≥ 10. Error bars represent standard error of the mean (SEM) from multiple repeats.

Immunostaining images confirmed typical morphological changes of an EMT from HGF treatment (**Fig. 4A**). In agreement with prior reports (*48, 49*), we observed down regulation of E-cadherin and up regulation of vimentin expression. We also observed differences in cell packing and cell morphology: untreated MDCK cells displayed a cobble stone morphology as a closely packed cell monolayer with E-cadherin staining colocalizing at the areas of lateral cell−cell contacts. In contrast, in the presence of HGF, the MDCK cells adopted a spindle-shaped morphology and vimentin expression was significantly higher. HGF-treated MDCK cells have lower value of the relaxation time constant ($\lambda = 9.6 \pm 0.9$ s$^{-1}$, $n = 3$, $N = 10$) than the untreated group ($\lambda = 12.2 \pm 1.2$ s$^{-1}$, $n = 3$, $N = 10$) (**Fig. 4C**). This indicates that HGF treated MDCK cell layer are softer than the untreated cells. This is expected for the following reasons: in untreated MDCK cells, polarized cells interact with the basement membrane through integrins and are held together through the adherent junctions (*50, 51*). During EMT, the overall monolayer stiffness decreases due to weakened cell-cell adhesions and rearrangements in the cytoskeleton. Adherent junctions degrade and are replaced by proteins that provide greater junctional flexibility (*52*). This is shown by the diminished presence of E-cadherin localized to the periphery of the cells as punctuated dots because mesenchymal cells lack apico-basal polarity, which further shows morphological rearrangement upon EMT. Previous studies showed that the presence of vimentin networks reduced the effective cytoskeletal mesh size and decreased the viscoelastic relaxation time constant ($\lambda$) of the cytoskeleton, resulting in a softer cell layer (*53*).

As part of our controls for this experiment, we also obtained the viscoelastic evolution of untreated MDCK cells during growth and monolayer formation. The relaxation time constant increases from the baseline of PDMS on day 1 to that of a confluent MDCK cell monolayer on day 7, as expected (**Fig. 4C**).

To confirm that stiffness changes are due to differences in cell phenotype and not compromised cell metabolism, we performed Prestoblue assay (**Fig. S3**). The result showed that there was no significant difference in metabolic activity between cells cultured with and without HGF, indicating that the relaxation time constant change was only due to phenotypic changes. Cell detachment also had negligible contributions to the relaxation time constant and stiffness. EMT has been linked to a significant cytoskeleton remodeling that weakens cell-cell adhesions but strengthens cell-matrix adhesions (*53*). Unlike MDCK cells treated with Trypsin-EDTA solution, HGF treated MDCK cells are more resistive to the shear force from fluid flow and are unlikely to detach from the PDMS channel wall.

## DISCUSSION

In this study, we developed a microfluidic system that leverages elastohydrodynamic phenomena to measure the viscoelasticity of cellular monolayers at both short (~minutes) and long (~days) time scale. EHD phenomena has a higher sensitivity compared to other techniques, likely due to the Boussinesq-like dependence of substrate deformation on additional layers (here, represented by the cellular monolayer). As the sensor is placed near the interface, at a distance of similar length scale to the cellular monolayers, small changes in monolayer size or mechanical properties lead to large changes in the substrate and sensor response. By using EHD, we obtained a baseline on MDCK cells with a 30-fold reduction in the noise. This improved signal-to-noise ratio made it possible to make direct mechanical measurements of monolayer integrity (trypsinization) and morphological changes (epithelial-to-mesenchymal transition). The short-term trypsinization experiments showed that both the cell body and extracellular matrix proteins contribute to the mechanical integrity of the monolayer, as expected. The long-term epithelial-to-mesenchymal transition experiments validated that epithelial monolayers become softer after EMT due to weakened intercellular adhesion. Throughout both experiments, the platform required minimal adjustments to cell culturing protocols and had significantly shorter acquisition times compared to AFM and other optical-based techniques, making it easier to maintain sterile technique. In principle, that platform now enables simultaneous labeling and long-time monitoring of viscoelasticity in the field of drug screening and disease pathogenesis. While we focused on monolayers here, we expect that the platform can be extended to thin films, including cells encapsulated in synthetic matrices. Given the wide use of cellular monolayers and importance of intercellular adhesions in diseases ranging fibrosis, sepsis, cancer metastasis, among others, this platform will enable rapid monitoring of phenotypic changes in viscoelasticity for use in both basic biology and drug discovery.

## MATERIALS AND METHODS

### Microfluidic platform design and fabrication.

Standard soft lithographic techniques were used to fabricate microfluidic devices (**Fig. 5**). Polycarbonate stock (75 mm × 50 mm × 9 mm, TAP Plastics) was cut using a computer numerical control (CNC) endmill (MDX-50 Desktop Milling Machine, Roland Corp.) to form a negative mold of the device (**Fig. 5A**). The mold was spin coated with 8% (w/w) poly(vinyl alcohol) (PVA, $M_w$ = 23,000 g/mol, Sigma-Aldrich Inc.) water solution at 300 rpm for 30 s after rinsed by ethanol and acetone and was dried on a hotplate. Then, PDMS (poly(dimethylsiloxane), Sylgard-184, Dow Corning Corp.) base and curing agent were mixed at a ratio of 20 : 1 or 15 : 1 by weight in a plastic cup, stirred thoroughly, and then degassed for 30 min. The PDMS was then poured onto the mold, scraped with a doctor blade to achieve a thin layer above the channel, and cured at 65 °C for 15 min (**Fig. 5B**). After curing, a rectangular piece of metallic nanoisland graphene strain sensor (~12 mm × 4 mm) with a copper backing layer was placed onto the negative channel spanning the short axis of the channel. After sensor placement, the copper backing was wet etched with a 15% w/w ammonium persulfate (APS) solution overnight at room temperature. After etching, the etchant was removed (**Fig. 5C**). A drop of eutectic gallium-indium (EGaIn) was placed on each side of the strain sensor to form electrical interconnects and then copper wires were placed into the EGaIn to form electrical leads. More PDMS was poured onto the cured PDMS (with the embedded graphene strain sensor) to the depth of 15 mm and was cured at 65 °C for 45 min (**Fig. 5D**). After curing, the device and mold were placed into warm water for 48 h in order to dissolve the PVA sacrificial release layer and release the PDMS block from the mold. After the PDMS block was separated from the mold and excess PDMS was trimmed from the device, fluid inlet and outlet holes were punched in the device using syringe needles and PTFE tubes were inserted. A microscope slide (75 mm × 50 mm, Thermo Scientific) was spun coated with PDMS with the same monomer-to-curing agent ratio at 500 rpm for 30 s to form a layer of 50 μm and then cured at 65°C for 7 min. The PDMS block was placed onto the glass slide coated with a partially cured layer of PDMS and was then cured for 30 min more at 65°C to seal the channel (**Fig. 5E**). After curing, the device was ready for testing. The dimensions of channel are 30 mm (*L*) × 1 mm (*W*) × 0.5 mm (*H*), and the sensor position ranges from 8 μm to 10 μm from the channel edge, as measured by a laser displacement sensor (KEYENCE LT-9010M, resolution 0.01μm).

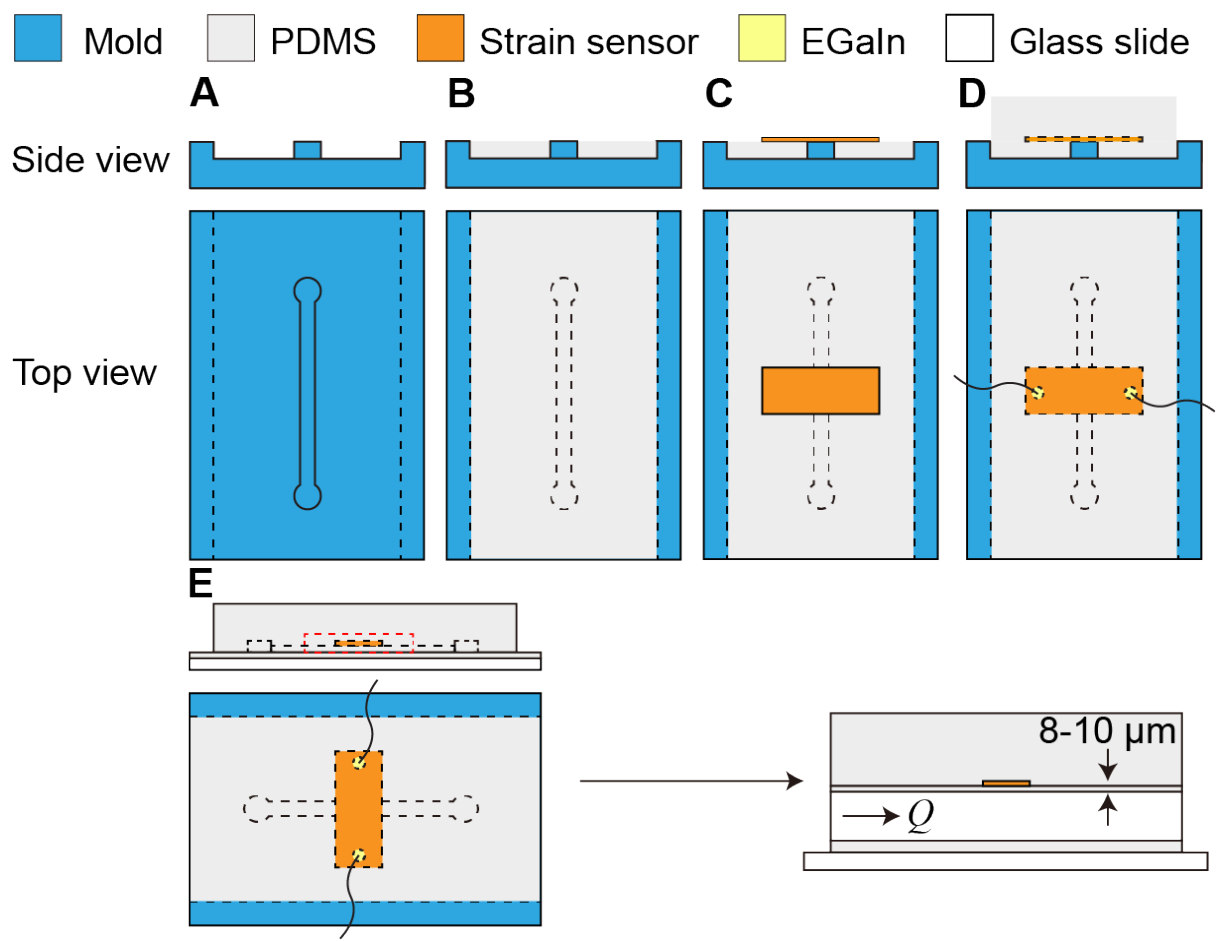

**Fig. 5. Microfluidic device fabrication.** Schematics of soft lithographic fabrication process. (**A**) Polycarbonate mold with a negative channel. (**B**) The first layer of PDMS was formed on the mold. (**C**) The graphene strain sensor was placed on the first layer of PDMS. (**D**) Eutectic

gallium-indium (EGaIn) was placed on each side of the strain sensor to form electrical interconnects and then copper wires were placed into the EGaIn to form electrical leads. The second layer of PDMS was formed on to encapsulate the sensor. (**E**) The device was bonded to the glass slide to seal the channel. The thickness of the PDMS layer between channel and strain sensor is ~8-10 μm.

**Metallic nanoisland-on-graphene fabrication.**
Palladium was evaporated on the single layer of graphene at rate of 0.04 Å/s. The strain sensor consists of a layer of copper foil, a layer of graphene and a layer of 8 nm palladium which provides high sensitivity at low strains and be able to transduce deformation to electrical signal. The mechanism for this high sensitivity is postulated to be a combination of crack formation, improved sensor integrity, and quantum tunneling. (*25*)

**Madin-Darby canine kidney (MDCK) cells culture and characterization.**
Immediately following device fabrication, all devices were sterilized in 70% ethanol for 20 min, followed by a 15-min exposure to a bactericidal UV lamp. Prior to cell culture, to promote cell attachment on PDMS, the PDMS channel wall was coated with collagen IV (collagen from human placenta, Sigma-Aldrich) (**Fig. 6A**). The device was incubated with 50 μg/mL collagen IV in 0.1% glacial acetic acid solution (in sterile PBS) at 37 °C in a humidified $CO_2$ incubator for 4 h and then washed with sterile PBS. The device was placed upside down to help the collagen coating process to the top of the channel, which was where the graphene sensor was located. MDCK cells were sub-cultured on two 75 $cm^2$ tissue culture-treated flasks. Cells were cultured and passaged according to manufacturer's protocol in Eagles's minimum essential medium (EMEM, ATCC) supplemented with 10% fetal bovine serum (FBS) and 100 IU/mL penicillin-streptomycin in a humidified 37°C incubator maintained at 5% $CO_2$ and 95% air. Cell culture media was changed every other day. Upon reaching confluency, cells were detached with 0.25% (w/v) trypsin-EDTA solution for 10 min and centrifuged at 450 g for 5 min. Cell pellets were re-suspended in EMEM media at a final concentration of 2 million cells/mL and the cell suspension was evenly pipetted into the channel. Cell culture media was refreshed every day. Just as with the collagen coating process, the device was placed upside down to promote cell attachment to the top of the channel wall, near the sensor.

**Contact angle measurement.**
Due to the size constraints of the channel structure, a separate substrate was coated with PDMS for the contact angle measurements. The PDMS solution with same monomer-to-curing agent ratio was spun-coated on a glass slide at 500 rpm for 30 seconds to form a PDMS film. The cured PDMS film was subjected to the same collagen IV modification described above. Contact angle hysteresis was measured at room temperature (**Fig. 6A**). A distilled water droplet was placed on the test substrates and the measured angles were averaged (3 samples of plain PDMS, 3 samples of collagen IV coated PDMS). Plain PDMS had a contact angle of (99.1° - 120.7°) ± 6.1°, which decreased to (84.4° - 103.7°) ± 5.4° after coating with collagen IV. The result shows collagen IV coating was successful.

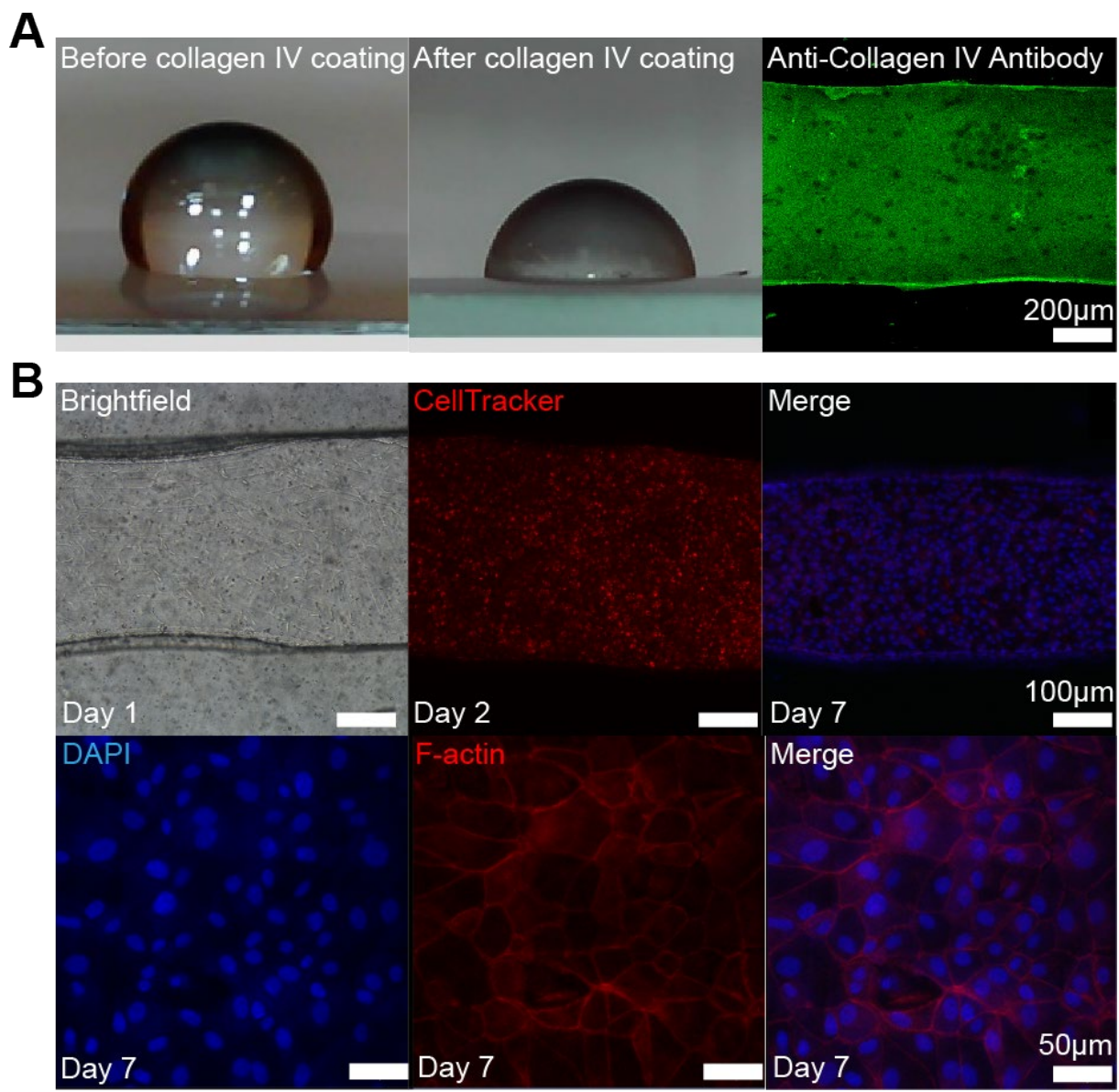

**Fig. 6. Collagen IV coating on PDMS and MDCK cell culture inside the microchannel.** (**A**) PDMS wettability changes after collagen IV coating which is shown by water contact angle measurement. The contact angle hysteresis shows contact angle of PDMS decreases from (99.1° - 120.7°) ± 6.1° to (84.4° - 103.7°) ± 5.4° after collagen IV coating. The immunostaining image of collagen IV (stained by monoclonal Anti-Collagen, Type IV antibody produced in mouse, Sigma-Aldrich) inside the microchannel confirms that the wall is coated with collagen IV. (**B**) Brightfield and confocal images of MDCK cells inside the microchannel. On day 7, MDCK cells become confluent inside the microchannel.

**Immunofluorescence.**

MDCK cells were fixed with 4% (w/v) paraformaldehyde (PFA, Sigma-Aldrich) in PBS for 15 min and followed by 3% BSA blocking incubation for nonspecific binding for 1 h at room temperature. Cells were then incubated with primary antibodies vinculin (Sigma-Aldrich) and E-cadherin (BD Biosciences) diluted in 3% BSA (1: 400) at room temperature for 4 h. The primary antibody solution was then aspirated, and the cells were washed 3 times by PBST solution. Secondary antibody Alexa Fluor 488 goat anti-mouse and Alexa Fluor 568 Phalloidin (Thermo Fisher Scientific) were diluted in 3% BSA (1 : 400 and 1 : 800, respectively) at room temperature for 4 h. After another triple wash with PBST solution, 4′,6-diamidino-2-phenylindole (DAPI, Thermo Fisher Scientific) was diluted in PBS solution (1 : 1000) at room temperature for 15 min. Fluorescence microscopy was performed using a Zeiss LSM 880 (**Fig. 6B**).

**Electrical signal capture of wall deformation via droplet flow and analysis.**
Viscoelastic behavior was modeled with a Kelvin-Voigt model which uses a combination of springs and dashpots connected in parallel. If a constant stress is applied to Kelvin-Voigt material and then released at time $t_1$, the viscoelastic relaxation obeys the following equation:

$$\varepsilon(t > t_1) = \varepsilon(t_1)e^{-\frac{t}{\tau}} \qquad (4)$$

where $E$ is the modulus of elasticity, $\eta$ is the viscosity and $\tau = \frac{\eta}{E}$ is the relaxation time.

To monitor this viscoelastic relaxation behavior, a brief stress is applied by flowing fluid droplet in the channel. As the sensor signal is proportional to wall deformation, the voltage change can be described in terms of the viscoelastic relaxation as:

$$V(t) = V_0 - V(t_1)e^{-\lambda t} \qquad (5)$$

where $\lambda = \frac{1}{\tau} = \frac{E}{\eta}$ is the relaxation time constant. To fit the viscoelastic relaxation, the equation is linearized as:

$$V(t) = V_0 - V(t_1)e^{-\lambda t} \rightarrow \ln\left(1 - \frac{V(t)}{V_0}\right) = -\lambda t + \ln\left(\frac{V(t_1)}{V_0}\right) \qquad (6)$$

where the slope of this linear function is the relaxation time constant $\lambda$.

**REFERENCES**

1. S. Suresh, Biomechanics and biophysics of cancer cells. *Acta Mater.* **55**, 3989–4014 (2007).
2. M. T. Frassica, M. A. Grunlan, Perspectives on synthetic materials to guide tissue regeneration for osteochondral defect repair. *ACS Biomater. Sci. Eng.* **6**, 4324–4336 (2020).
3. C. A. Reinhart-King, M. Dembo, D. A. Hammer, Cell-cell mechanical communication through compliant substrates. *Biophys. J.* **95**, 6044–6051 (2008).
4. D. A. Fletcher, R. D. Mullins, Cell mechanics and the cytoskeleton. *Nature.* **463**, 485–492 (2010).

5.  C. Guillot, T. Lecuit, Mechanics of epithelial tissue homeostasis and morphogenesis. *Science (80-. )*. **340**, 1185–1189 (2013).
6.  J. M. Barnes, L. Przybyla, V. M. Weaver, Tissue mechanics regulate brain development, homeostasis and disease. *J. Cell Sci.* **130**, 71–82 (2017).
7.  Y. Nematbakhsh, C. T. Lim, Cell biomechanics and its applications in human disease diagnosis. *Acta Mech. Sin.* **31**, 268–273 (2015).
8.  J. S. Saini, B. Corneo, J. D. Miller, T. R. Kiehl, Q. Wang, N. C. Boles, T. A. Blenkinsop, J. H. Stern, S. Temple, Nicotinamide ameliorates disease phenotypes in a human iPSC model of age-related macular degeneration. *Cell Stem Cell*. **20**, 635–647 (2017).
9.  M. Deiana, S. Calfapietra, A. Incani, A. Atzeri, D. Rossin, R. Loi, B. Sottero, N. Iaia, G. Poli, F. Biasi, Derangement of intestinal epithelial cell monolayer by dietary cholesterol oxidation products. *Free Radic. Biol. Med.* **113**, 539–550 (2017).
10. S. L. Friedman, D. Sheppard, J. S. Duffield, S. Violette, Therapy for fibrotic diseases: Nearing the starting line. *Sci. Transl. Med.* **5** (2013), doi:10.1126/scitranslmed.3004700.
11. Y. M. Efremov, I. M. Zurina, V. S. Presniakova, N. V Kosheleva, D. V Butnaru, A. A. Svistunov, Y. A. Rochev, P. S. Timashev, Mechanical properties of cell sheets and spheroids: the link between single cells and complex tissues. *Biophys. Rev.* **13**, 541–561 (2021).
12. L. Aoun, S. Larnier, P. Weiss, M. Cazales, A. Herbulot, B. Ducommun, C. Vieu, V. Lobjois, Measure and characterization of the forces exerted by growing multicellular spheroids using microdevice arrays. *PLoS One*. **14**, e0217227 (2019).
13. J. C. Villalobos Lizardi, J. Baranger, M. B. Nguyen, A. Asnacios, A. Malik, J. Lumens, L. Mertens, M. K. Friedberg, C. A. Simmons, M. Pernot, A guide for assessment of myocardial stiffness in health and disease. *Nat. Cardiovasc. Res.* **1**, 8–22 (2022).
14. A. Fuhrmann, J. R. Staunton, V. Nandakumar, N. Banyai, P. C. W. Davies, R. Ros, AFM stiffness nanotomography of normal, metaplastic and dysplastic human esophageal cells. *Phys. Biol.* **8**, 15007 (2011).
15. M. Dao, C. T. Lim, S. Suresh, Mechanics of the human red blood cell deformed by optical tweezers. *J. Mech. Phys. Solids*. **51**, 2259–2280 (2003).
16. F. K. Glenister, R. L. Coppel, A. F. Cowman, N. Mohandas, B. M. Cooke, Contribution of parasite proteins to altered mechanical properties of malaria-infected red blood cells. *Blood*. **99**, 1060–1063 (2002).
17. B. Srinivasan, A. R. Kolli, M. B. Esch, H. E. Abaci, M. L. Shuler, J. J. Hickman, TEER measurement techniques for in vitro barrier model systems. *SLAS Technol.* **20**, 107–126 (2015).
18. P. Polimeno, A. Magazzu, M. A. Iati, F. Patti, R. Saija, C. D. E. Boschi, M. G. Donato, P. G. Gucciardi, P. H. Jones, G. Volpe, Optical tweezers and their applications. *J. Quant. Spectrosc. Radiat. Transf.* **218**, 131–150 (2018).
19. R. W. Style, R. Boltyanskiy, G. K. German, C. Hyland, C. W. MacMinn, A. F. Mertz,

L. A. Wilen, Y. Xu, E. R. Dufresne, Traction force microscopy in physics and biology. *Soft Matter*. **10**, 4047–4055 (2014).
20. F. Li, J. H.-C. Wang, Q.-M. Wang, Thickness shear mode acoustic wave sensors for characterizing the viscoelastic properties of cell monolayer. *Sensors Actuators B Chem.* **128**, 399–406 (2008).
21. A. Adamo, A. Sharei, L. Adamo, B. Lee, S. Mao, K. F. Jensen, Microfluidics-based assessment of cell deformability. *Anal. Chem.* **84**, 6438–6443 (2012).
22. J. Guck, S. Schinkinger, B. Lincoln, F. Wottawah, S. Ebert, M. Romeyke, D. Lenz, H. M. Erickson, R. Ananthakrishnan, D. Mitchell, J. Kas, S. Ulvick, C. Bilby, Optical deformability as an inherent cell marker for testing malignant transformation and metastatic competence. *Biophys. J.* **88**, 3689–3698 (2005).
23. Y. Man, E. Kucukal, R. An, Q. D. Watson, J. Bosch, P. A. Zimmerman, J. A. Little, U. A. Gurkan, Microfluidic assessment of red blood cell mediated microvascular occlusion. *Lab Chip*. **20**, 2086–2099 (2020).
24. P. M. Holloway, S. Willaime-Morawek, R. Siow, M. Barber, R. M. Owens, A. D. Sharma, W. Rowan, E. Hill, M. Zagnoni, Advances in microfluidic in vitro systems for neurological disease modeling. *J. Neurosci. Res.* **99**, 1276–1307 (2021).
25. A. V. Zaretski, S. E. Root, A. Savchenko, E. Molokanova, A. D. Printz, L. Jibril, G. Arya, M. Mercola, D. J. Lipomi, Metallic Nanoislands on Graphene as Highly Sensitive Transducers of Mechanical, Biological, and Optical Signals. *Nano Lett.* **16**, 1375–1380 (2016).
26. X. Wang, I. C. Christov, Theory of the flow-induced deformation of shallow compliant microchannels with thick walls. *Proc. R. Soc. A*. **475**, 20190513 (2019).
27. I. C. Christov, V. Cognet, T. C. Shidhore, H. A. Stone, Flow rate-pressure drop relation for deformable shallow microfluidic channels. *J. Fluid Mech.* **841**, 267–286 (2018).
28. E. Boyko, H. A. Stone, I. C. Christov, Flow rate-pressure drop relation for deformable channels via fluidic and elastic reciprocal theorems. *Phys. Rev. Fluids*. **7**, L092201 (2022).
29. G. Guyard, F. Restagno, J. D. Mcgraw, Elastohydrodynamic relaxation of soft and deformable microchannels. *arXiv Prepr. arXiv2207.05408* (2022).
30. T. Gervais, J. El-Ali, A. Günther, K. F. Jensen, Flow-induced deformation of shallow microfluidic channels. *Lab Chip*. **6**, 500–507 (2006).
31. C. Dhong, S. J. Edmunds, J. Ramírez, L. V Kayser, F. Chen, J. V Jokerst, D. J. Lipomi, Optics-Free, Non-Contact Measurements of Fluids, Bubbles, and Particles in Microchannels Using Metallic Nano-Islands on Graphene. *Nano Lett.* **18**, 5306–5311 (2018).
32. A. Hemmasizadeh, M. Autieri, K. Darvish, Multilayer material properties of aorta determined from nanoindentation tests. *J. Mech. Behav. Biomed. Mater.* **15**, 199–207 (2012).
33. M. Malakouti, M. Ameri, P. Malekzadeh, Dynamic viscoelastic incremental-layerwise finite element method for multilayered structure analysis based on the relaxation

approach. *J. Mech.* **30**, 593–602 (2014).
34. H. S. Gupta, J. Seto, S. Krauss, P. Boesecke, H. R. C. Screen, In situ multi-level analysis of viscoelastic deformation mechanisms in tendon collagen. *J. Struct. Biol.* **169**, 183–191 (2010).
35. N. Alam, N. T. Asnani, Vibration and damping analysis of multilayered rectangular plates with constrained viscoelastic layers. *J. Sound Vib.* **97**, 597–614 (1984).
36. S. R. Aglyamov, S. Wang, A. B. Karpiouk, J. Li, M. Twa, S. Y. Emelianov, K. V Larin, The dynamic deformation of a layered viscoelastic medium under surface excitation. *Phys. Med. Biol.* **60**, 4295–4312 (2015).
37. Y. Huang, K. Choi, C. H. Hidrovo, The improved resistance of PDMS to pressure-induced deformation and chemical solvent swelling for microfluidic devices. *Microelectron. Eng.* **124**, 66–75 (2014).
38. D. P. Gaver, D. Halpern, O. E. Jensen, J. B. Grotberg, The steady motion of a semi-infinite bubble through a flexible-walled channel. *J. Fluid Mech.* **319**, 25–65 (1996).
39. J. E. Wagenseil, R. P. Mecham, Vascular extracellular matrix and arterial mechanics. *Physiol. Rev.* **89**, 957–989 (2009).
40. N. Khalilgharibi, J. Fouchard, N. Asadipour, R. Barrientos, M. Duda, A. Bonfanti, A. Yonis, A. Harris, P. Mosaffa, Y. Fujita, Stress relaxation in epithelial monolayers is controlled by the actomyosin cortex. *Nat. Phys.* **15**, 839–847 (2019).
41. B. L. Schumann, T. E. Cody, M. L. Miller, G. D. Leikauf, Isolation, characterization, and long-term culture of fetal bovine tracheal epithelial cells. *Vitr. Cell. Dev. Biol.* **24**, 211–216 (1988).
42. F. Sorba, A. Poulin, R. Ischer, H. Shea, C. Martin-Olmos, Integrated elastomer-based device for measuring the mechanics of adherent cell monolayers. *Lab Chip*. **19**, 2138–2146 (2019).
43. C. M. Flanigan, K. R. Shull, Adhesive and elastic properties of thin gel layers. *Langmuir*. **15**, 4966–4974 (1999).
44. R. Strippoli, R. Moreno-Vicente, C. Battistelli, C. Cicchini, V. Noce, L. Amicone, A. Marchetti, M. A. Del Pozo, M. Tripodi, Molecular mechanisms underlying peritoneal EMT and fibrosis. *Stem Cells Int.* **2016** (2016).
45. R. Kalluri, R. A. Weinberg, The basics of epithelial-mesenchymal transition. *J. Clin. Invest.* **119**, 1420–1428 (2009).
46. F. Liu, S. Song, Z. Yi, M. Zhang, J. Li, F. Yang, H. Yin, X. Yu, C. Guan, Y. Liu, HGF induces EMT in non-small-cell lung cancer through the hBVR pathway. *Eur. J. Pharmacol.* **811**, 180–190 (2017).
47. A. Ravikrishnan, T. Ozdemir, M. Bah, K. A. Baskerville, S. I. Shah, A. K. Rajasekaran, X. Jia, Regulation of epithelial-to-mesenchymal transition using biomimetic fibrous scaffolds. *ACS Appl. Mater. Interfaces*. **8**, 17915–17926 (2016).
48. M. K. Wendt, M. A. Taylor, B. J. Schiemann, W. P. Schiemann, Down-regulation of epithelial cadherin is required to initiate metastatic outgrowth of breast cancer. *Mol. Biol. Cell*. **22**, 2423–2435 (2011).
49. S. Wu, Y. Du, J. Beckford, H. Alachkar, Upregulation of the EMT marker vimentin is

associated with poor clinical outcome in acute myeloid leukemia. *J. Transl. Med.* **16**, 1–9 (2018).
50. H. Oda, K. Tagawa, Y. Akiyama-Oda, Diversification of epithelial adherens junctions with independent reductive changes in cadherin form: identification of potential molecular synapomorphies among bilaterians. *Evol. Dev.* **7**, 376–389 (2005).
51. P. Hulpiau, F. Van Roy, Molecular evolution of the cadherin superfamily. *Int. J. Biochem. Cell Biol.* **41**, 349–369 (2009).
52. M. S. Balda, K. Matter, Tight junctions at a glance. *J. Cell Sci.* **121**, 3677–3682 (2008).
53. S. E. Leggett, A. M. Hruska, M. Guo, I. Y. Wong, The epithelial-mesenchymal transition and the cytoskeleton in bioengineered systems. *Cell Commun. Signal.* **19**, 32 (2021).

**Acknowledgements**
We would like to thank Dr. Yong Zhao for the assistance with the scanning electron microscope. We acknowledge the W. M. Keck Center for the use of the scanning electron microscopy and the Bio-imaging Center of the Delaware Biotechnology Institute for the use of the confocal microscope.

**Funding:**
National Institutes of Health grant R01DC014461 (XJ, CD, TG, XZ)
National Institutes of Health grant P20GM139760 (CD, SS)
University of Delaware Gore Graduate Fellowship (SS)

**Author contributions:**
Conceptualization: TG, XJ, CD
Methodology: TG, XZ, SS
Investigation: TG, XZ
Supervision: XJ, CD
Writing-original draft: TZ, CD
Writing-review & editing: TZ, XZ, SS, XJ, CD

**Competing interests:**
Authors declare that they have no competing interests.

**Data and materials availability:**
All data are available in the main text or the supplementary materials.